\documentclass [12 pt]{article}
\def\be{\begin{equation}}
\def\eea{\end{eqnarray}}
\def\bea{\begin{eqnarray}}
\def\ee{\end{equation}}

\usepackage[psamsfonts]{amssymb}
\usepackage{amsmath}

\author{M. Amooshahi$^{1}$ \footnote{amooshahi@sci.ui.ac.ir}
\\ $^{1}$ {\small Faculty of science, University of Isfahan ,Hezar Jarib Ave.,
Isfahan,Iran}}
\title{Canonical quantization of electromagnetic field  in  the presence of nonlinear anisotropic
magnetodielectric medium with spatial-temporal dispersion}
\begin{document}
\maketitle
\begin{abstract}
Modeling a nonlinear anisotropic magnetodielectric medium with
spatial-temporal dispersion by two continuum collections of three
dimensional harmonic oscillators, a fully canonical quantization of
the electromagnetic field is demonstrated in the presence of such a
medium. Some coupling tensors of various ranks are introduced that
couple the magnetodielectric medium with the electromagnetic field.
The polarization and magnetization fields of the medium are defined
in terms of the coupling tensors and the oscillators modeling the
medium. The electric and magnetic susceptibility tensors of the
medium are obtained in terms of the coupling tensors. It is shown
that the electric field satisfy an integral equation in frequency
domain. The integral equation is solved by an iteration method and
the electric field is found up to an arbitrary accuracy.
\end{abstract}
\begin{description}
\item[PACS number] 12.20.Ds
\end{description}
\section{Introduction}
The quantum features of the electromagnetic field can be influenced
by the magnetodielectric media. For example the Casimir effect
\cite{1} - \cite{10} and the spontaneous emission rate of an
initially excited atom \cite{11}-\cite{14} are modified in the
presence of magnetodielectric media. There is a canonical
quantization approach for the macroscopic electromagnetic field in a
linear absorbing dielectric based on the damped polarization
model\cite{15}. In this scheme the electric polarization field of
the medium is appeared in the Lagrangian of the total system as a
part of the degrees of freedom of the medium. The other parts of the
degrees of freedom of the absorbing dielectric are related to the
dynamical variables of a reservoir coupled to the polarization field
in order to inclusion the absorption property of the medium. The
Hamiltonian of the total system is diagonalized in two steps. The
first step is diagonalization of the polarization-reservoir part and
the second step is diagonalization of the total Hamiltonian. This
method has
been generalized to inhomogeneous dielectrics \cite{16}and anisotropic magnetodielectric media \cite{17}.\\
There is a new fully canonical quantization of the electromagnetic
field in the presence of linear absorbing anisotropic
magnetodielectric media in which the medium is modeled by two
continuum collections of three dimensional space dependent harmonic
oscillators\cite{18},\cite{19}. One of the collection describes the
electric property  and the another collection explains the magnetic
property of the magnetodielectric medium. In contrast to the damped
polarization model it is not needed the electric and magnetic
polarization fields to be appeared in the Lagrangian of the total
system. The two collections of the harmonic oscillators solely
constitute the degrees of freedom of the medium. In fact the space
dependent harmonic oscillators  are able to describe both absorption
and polarization properties of the medium.
 This means that in this approach the electric and magnetic polarization fields  are
defined in terms of the harmonic oscillators modeling the medium and
the coupling tensors which couple   the medium and electromagnetic
field. Among the advantages of this approach is that this approach
dose not need the processes  of the diagonalization of the
Hamiltonians of the different parts of the total system and  is
simply generalizable to the nonlinear anisotropic magnetodielectric
medium with spatial-temporal dispersion. The aim of the present work
is generalization of the mentioned scheme for a nonlinear
anisotropic magnetodielectric medium with spatial-temporal
dispersion. The canonical quantization of a three dimensional system
in the presence of a nonlinear anisotropic absorbing environment has
been done previously and the effect of nonlinearity of the
environment on the spontaneous emission of an initially excited two
level atom has been investigated\cite{20}
\section{The Lagrangian of the total system}
Modeling  the nonlinear anisotropic  magnetodielectric medium by two
continuum collections three dimensional harmonic oscillators, the
Lagrangian of the total system, that is the electromagnetic field
plus the magnetodielectric medium, is introduced as the sum of three
parts
\begin{equation}\label{n1}
 L(t)=  L_{em}(t)+L_{res}(t)+L_{int}(t)
 \end{equation}
 The first part is the Lagrangian of the electromagnetic field which
 as usual can be written as
\begin{equation}\label{n2}
L_{em}(t)=\int d^3r\left[\frac{1}{2}\varepsilon_0 {\bf E}^2({\bf
r},t)+\frac{{\bf B}^2({\bf r},t)}{2\mu_0}\right]
\end{equation}
where ${\bf E}=-\vec{\nabla \varphi}-\frac{\partial {\bf
A}}{\partial t}$ and ${\bf B}=\nabla\times {\bf A}$ and ${\bf A} ,
\varphi $ are respectively the vector and scalar potential which
constitute the dynamical variables related to of the electromagnetic
field. The second part in (\ref{n1}) is the Lagrangian of the
magnetodielectric medium. If we denote the two continuum collections
of three dimensional harmonic oscillators modeling the medium by $
{\bf X}_\omega({\bf r},t)$ and $ {\bf Y}_\omega({\bf r},t)$, with
continuous parameter $\omega $, then the Lagrangian of the medium is
written as
\begin{eqnarray}\label{n3}
L_{res}(t)&=&\int d^3r \int_0^\infty d\omega
\left[\frac{1}{2}\dot{{\bf X}}_\omega^2({\bf
r},t)-\frac{1}{2}\omega^2{\bf
X}^2_\omega({\bf r},t)\right]\nonumber\\
&+&\int d^3r \int_0^\infty d\omega \left[\frac{1}{2}\dot{{\bf
Y}}^2_\omega({\bf r},t)-\frac{1}{2}\omega^2{\bf Y}^2_\omega({\bf
r},t)\right]
\end{eqnarray}
It will be shown that the two collections ${\bf X}_\omega({\bf
r},t)$ and $ {\bf Y}_\omega({\bf r},t)$ describe, respectively, the
electric and magnetic properties of the medium. That is the electric
polarization field of the medium is defined in terms of ${\bf
X}_\omega({\bf r},t)$ and the magnetic polarization field of the
medium is expressed in terms of ${\bf Y}_\omega({\bf r},t)$.\\
The third term in (\ref{n1}) is the interaction part of the
electromagnetic field and the magnetodielectric medium that for a
nonlinear anisotropic medium with spatial-temporal dispersion is
proposed as
\begin{eqnarray}\label{n4}
&&L_{int}(t)=\int_0^\infty d\omega_1\int d^3r \int d^3 r_1
f^{(1)}_{i i_1}(\omega_1, {\bf r}, {\bf r}_1) E^i({\bf r},t)
X^{i_1}_{\omega_1}({\bf r}_1,t)\nonumber\\
&&+\int_0^\infty d\omega_1 \int_0^\infty d\omega_2\int d^3r \int d^3
r_1 \int d^3r_2[f^{(2)}_{i i_1 i_2}(\omega_1, \omega_2,{\bf r}, {\bf
r}_1, {\bf r}_2)\nonumber\\
&&\times E^i({\bf r},t) X^{i_1}_{\omega_1}({\bf
r}_1,t)X^{i_2}_{\omega_2}({\bf r}_2,t)]\nonumber\\
&&+\int_0^\infty d\omega_1 \int_0^\infty d\omega_2 \int_0^\infty d
\omega_3\int d^3r \int d^3 r_1 \int d^3r_2 \int d^3r_3 [f^{(3)}_{i
i_1 i_2 i_3}(\omega_1, \omega_2,\omega_3,{\bf r}, {\bf r}_1, {\bf
r}_2, {\bf r}_3) \nonumber\\
&&\times E^i({\bf r},t) X^{i_1}_{\omega_1}({\bf
r}_1,t)X^{i_2}_{\omega_2}({\bf r}_2,t)X^{i_3}_{\omega_3}({\bf r}_3,t)]+...\nonumber\\
&&+\int_0^\infty d\omega_1\int d^3r \int d^3 r_1 g^{(1)}_{i
i_1}(\omega_1, {\bf r}, {\bf r}_1) B^i({\bf r},t)
Y^{i_1}_{\omega_1}({\bf r}_1,t)\nonumber\\
&&+\int_0^\infty d\omega_1 \int_0^\infty d\omega_2\int d^3r \int d^3
r_1 \int d^3r_2[g^{(2)}_{i i_1 i_2}(\omega_1, \omega_2,{\bf r}, {\bf
r}_1, {\bf r}_2)\nonumber\\
&&\times B^i({\bf r},t) Y^{i_1}_{\omega_1}({\bf
r}_1,t)Y^{i_2}_{\omega_2}({\bf r}_2,t)]\nonumber\\
&&+\int_0^\infty d\omega_1 \int_0^\infty d\omega_2 \int_0^\infty d
\omega_3\int d^3r \int d^3 r_1 \int d^3r_2 \int d^3r_3 [g^{(3)}_{i
i_1 i_2 i_3}(\omega_1, \omega_2,\omega_3,{\bf r}, {\bf r}_1, {\bf
r}_2, {\bf r}_3) \nonumber\\
&&\times B^i({\bf r},t) Y^{i_1}_{\omega_1}({\bf
r}_1,t)Y^{i_2}_{\omega_2}({\bf r}_2,t)Y^{i_3}_{\omega_3}({\bf r}_3,t)]+...\nonumber\\
\end{eqnarray}
where the tensors ${\bf f}^{(1)}, {\bf f}^{(2)}, {\bf f}^{(3)}, ...
$ and ${\bf g}^{(1)}, {\bf g}^{(2)}, {\bf g}^{(3)}, ...$ are the
coupling tensors between the electromagnetic field and the nonlinear
magnetodielectric medium. These coupling tensors play an important
role in this quantization approach. It will be shown that the
electric polarization field is obtained in terms of the coupling
tensors ${\bf f}^{(1)}, {\bf f}^{(2)}, {\bf f}^{(3)}, ... $ and the
harmonic oscillators ${\bf X}_\omega({\bf r},t)$. Also the electric
susceptibilities tensors of the medium  are expressed in terms of
the coupling tensors ${\bf f}^{(1)}, {\bf f}^{(2)}, {\bf f}^{(3)},
... $. Therefore these tensors are an indication for the ability of
the polarization of the medium. Similarly the magnetic polarization
field of the magnetodielectric medium is defined in terms of the
coupling tensors ${\bf g}^{(1)}, {\bf g}^{(2)}, {\bf g}^{(3)}, ... $
and the harmonic oscillators ${\bf Y}_\omega({\bf r},t)$. As well
the magnetic susceptibility tensors of the medium are written in
terms of the coupling tensors ${\bf g}^{(1)}, {\bf g}^{(2)}, {\bf
g}^{(3)}, ... $ and therefore these tensors explain the ability of
magnetization of the medium. The Lagrangian (\ref{n1})-(\ref{n4}) is
a nonlocal one and there is some problems for a canonical
quantization using this Lagrangian. The easiest way to overcome the
difficulties related to a nonlocal Lagrangian is to work in the
reciprocal space. Therefore we write the classical fields and the
coupling tensors appeared in the Lagrangian (\ref{n1})-(\ref{n4}) in
terms of their spatial fourier transforms. For example for the field
${\bf X}_\omega({\bf r},t)$ we write
\begin{equation}\label{n5}
{\bf X}_\omega({\bf r},t)=\frac{1}{(2\pi)^{\frac{3}{2}}}\int d^3k
 {\bf \underline{X}}_\omega({\bf k},t)e^{i {\bf k}\cdot {\bf r}}
\end{equation}
Similarly the components of the coupling tensors ${\bf f}^{(n)}$ and
${\bf g}^{(n)}$ can be written in terms of their Fourier transforms
as
\begin{eqnarray}\label{n6}
&&f^{(n)}_{i i_1 i_2...i_n}(\omega_1,\omega_2,...\omega_n,{\bf
r},{\bf r}_1,...,{\bf
r}_n)\nonumber\\
&&=\frac{1}{\frac{3(n+1)}{2}}\int d^3k \int d^3k_1...\int d^3k_n
\underline{f}^{(n)}_{i i_1...i_n}(\omega_1,...,\omega_n,{\bf k},{\bf
k}_1,...,{\bf k}_n)e^{i {\bf k}\cdot{\bf r}-i{\bf k}_1\cdot{\bf
r}_1-...-i{\bf k}_n\cdot{\bf r}_n}\nonumber\\
&&g^{(n)}_{i i_1 i_2...i_n}(\omega_1,\omega_2,...\omega_n,{\bf
r},{\bf r}_1,...,{\bf
r}_n)\nonumber\\
&&=\frac{1}{\frac{3(n+1)}{2}}\int d^3k \int d^3k_1...\int d^3k_n
\underline{g}^{(n)}_{i i_1...i_n}(\omega_1,...,\omega_n,{\bf k},{\bf
k}_1,...,{\bf k}_n)e^{i {\bf k}\cdot{\bf r}-i{\bf k}_1\cdot{\bf
r}_1-...-i{\bf k}_n\cdot{\bf r}_n}\nonumber\\
&&
\end{eqnarray}
Since the fields  appeared in Lagrangian (\ref{n1})-(\ref{n4})are
real valued , we have
 \begin{equation}\label{n7}
 \underline{{\bf X}}_\omega({\bf
k},t)=\underline{{\bf X}}^*_\omega({\bf k},t)
\end{equation}
for the field ${\bf X}_\omega({\bf r},t)$
 and the other fields in the Lagrangian
(\ref{n1})-(\ref{n4}). Similar relations for the real valued
coupling tensors ${\bf f}^{(n)}$ and ${\bf g}^{(n)}$ are as follows
\begin{eqnarray}\label{n8}
\underline{f}^{(n)}_{i i_1...i_n}(\omega_1,...,\omega_n,{\bf k},{\bf
k}_1,...,{\bf k}_n)&=&(\underline{f}^{(n)})^*_{i
i_1...i_n}(\omega_1,...,\omega_n,-{\bf k},-{\bf k}_1,...,-{\bf
k}_n)\nonumber\\
\underline{g}^{(n)}_{i i_1...i_n}(\omega_1,...,\omega_n,{\bf k},{\bf
k}_1,...,{\bf k}_n)&=&(\underline{g}^{(n)})^*_{i
i_1...i_n}(\omega_1,...,\omega_n,-{\bf k},-{\bf k}_1,...,-{\bf
k}_n)\nonumber\\
&&
\end{eqnarray}
Because of the relations (\ref{n7}) and (\ref{n8}) the independent
dynamical variables can be taken into account by restricting the
integration domain to the half space $k_z\geq 0$ in the reciprocal
space. Substituting the real valued fields and the coupling tensors
${\bf f}^{(n)}$ and $ {\bf g}^{(n)}$ in the total Lagrangian
(\ref{n1})-(\ref{n4}) by their Fourier transforms and using the
relation (\ref{n7}) for ${\bf X}_\omega({\bf r},t)$, and similar
relations for the other fields and also using (\ref{n8}),one can
rewrite the Lagrangian of the total system as
\begin{eqnarray}\label{n9}
&&L(t)=\int' d^3k\left[\varepsilon_0| \underline{\dot{{\bf A}}}({\bf
k},t)|^2+\varepsilon_0 k^2|\underline{\varphi}({\bf
k},t)|^2-\frac{|{\bf k}\times\underline{{\bf A}}({\bf
k},t)|^2}{\mu_0}\right]\nonumber\\
&&+\varepsilon_0\int' d^3k\left[ -i {\bf k}\cdot
\underline{\dot{A}}({\bf k},t)\underline{\varphi}^*({\bf
k},t)+H.C\right] \nonumber\\
&&+\int_0^\infty d\omega\int' d^3k\left[|\underline{\dot{{\bf
X}}}_\omega({\bf k},t)|^2-\omega^2|\underline{{\bf X}}_\omega({\bf
k},t)|^2\right]\nonumber\\
&&+\int_0^\infty d\omega\int' d^3k\left[|\underline{\dot{{\bf
Y}}}_\omega({\bf k},t)|^2-\omega^2|\underline{{\bf Y}}_\omega({\bf
k},t)|^2\right]\nonumber\\
&&+\int_0^\infty d\omega_1\int' d^3k \int' d^3k_1
\left[\underline{f}^{(1)}_{ij}(\omega_1,{\bf k}, {\bf
k_1})\underline{E}^{*^i}({\bf k},t)\underline{X}^j_{\omega_1}({\bf
k_1},t)+H.C\right]\nonumber\\
&&+\int_0^\infty d\omega_1\int' d^3k \int' d^3k_1
\left[\underline{f}^{(1)}_{ij}(\omega_1,-{\bf k}, {\bf
k_1})\underline{E}^i({\bf k},t)\underline{X}^j_{\omega_1}({\bf
k_1},t)+H.C\right]\nonumber\\
&&+\int_0^\infty d\omega_1 \int_0^\infty d\omega_2\int' d^3k \int'
d^3k_1\int' d^3k_2 [\underline{f}^{(2)}_{ijk}(\omega_1,\omega_2,{\bf
k}, {\bf k}_1, {\bf k}_2) \nonumber\\
&&\times\underline{E}^{*^i}({\bf
k},t)\underline{X}^j_{\omega_1}({\bf
k}_1,t)\underline{X}^k_{\omega_2}({\bf k}_2,t)+H.C]\nonumber\\
&&+\int_0^\infty d\omega_1 \int_0^\infty d\omega_2\int' d^3k \int'
d^3k_1\int' d^3k_2[\underline{f}^{(2)}_{ijk}(\omega_1,\omega_2,{\bf
k}, {\bf k}_1, -{\bf k}_2)\nonumber\\
&&\times\underline{E}^{*^i}({\bf
k},t)\underline{X}^j_{\omega_1}({\bf
k}_1,t)\underline{X}^{*^k}_{\omega_2}({\bf k}_2,t)+H.C]\nonumber\\
&&+\int_0^\infty d\omega_1 \int_0^\infty d\omega_2\int' d^3k \int'
d^3k_1\int' d^3k_2 [\underline{f}^{(2)}_{ijk}(\omega_1,\omega_2,{\bf
k},- {\bf k}_1, {\bf k}_2)\nonumber\\
&&\times\underline{E}^{*^i}({\bf
k},t)\underline{X}^{*^j}_{\omega_1}({\bf
k}_1,t)\underline{X}^{k}_{\omega_2}({\bf k}_2,t)+H.C]\nonumber\\
&&+\int_0^\infty d\omega_1 \int_0^\infty d\omega_2\int' d^3k \int'
d^3k_1\int' d^3k_2 [\underline{f}^{(2)}_{ijk}(\omega_1,\omega_2,{\bf
k},- {\bf k}_1,- {\bf k}_2)\nonumber\\
&&\times\underline{E}^{*^i}({\bf
k},t)\underline{X}^{*^j}_{\omega_1}({\bf
k}_1,t)\underline{X}^{*^k}_{\omega_2}({\bf k}_2,t)+H.C]+....\nonumber\\
&&+\int_0^\infty d\omega_1\int'd^3k \int' d^3k_1
[\underline{g}^{(1)}_{ij}(\omega_1,{\bf k}, {\bf
k}_1)\underline{B}^{*^i}({\bf k},t)\underline{Y}^j_{\omega_1}({\bf
k}_1,t)+H.C]\nonumber\\
&&+\int_0^\infty d\omega_1\int'd^3k \int' d^3k_1
[\underline{g}^{(1)}_{ij}(\omega_1,-{\bf k}, {\bf
k}_1)\underline{B}^{i}({\bf k},t)\underline{Y}^j_{\omega_1}({\bf
k}_1,t)+H.C]\nonumber\\
&&+\int_0^\infty d\omega_1 \int_0^\infty d\omega_2\int' d^3k \int'
d^3k_1\int' d^3k_2[\underline{g}^{(2)}_{ijk}(\omega_1,\omega_2,{\bf
k}, {\bf k}_1, {\bf k}_2)\nonumber\\
&&\times\underline{B}^{*^i}({\bf
k},t)\underline{Y}^j_{\omega_1}({\bf
k}_1,t)\underline{Y}^{k}_{\omega_2}({\bf k}_2,t)+H.C]\nonumber\\
&&+\int_0^\infty d\omega_1 \int_0^\infty d\omega_2\int' d^3k \int'
d^3k_1\int' d^3k_2[\underline{g}^{(2)}_{ijk}(\omega_1,\omega_2,{\bf
k}, {\bf k}_1, -{\bf k}_2)\nonumber\\
&&\times\underline{B}^{*^i}({\bf
k},t)\underline{Y}^j_{\omega_1}({\bf
k}_1,t)\underline{Y}^{*^k}_{\omega_2}({\bf k}_2,t)+H.C]\nonumber\\
&&+\int_0^\infty d\omega_1 \int_0^\infty d\omega_2\int' d^3k \int'
d^3k_1\int' d^3k_2[\underline{g}^{(2)}_{ijk}(\omega_1,\omega_2,{\bf
k},- {\bf k}_1, {\bf k}_2)\nonumber\\
&&\times\underline{B}^{*^i}({\bf
k},t)\underline{Y}^{*^j}_{\omega_1}({\bf
k}_1,t)\underline{Y}^{k}_{\omega_2}({\bf k}_2,t)+ H.C]\nonumber\\
&&+\int_0^\infty d\omega_1 \int_0^\infty d\omega_2\int' d^3k \int'
d^3k_1\int' d^3k_2[\underline{g}^{(2)}_{ijk}(\omega_1,\omega_2,{\bf
k},- {\bf k}_1, -{\bf k}_2)\nonumber\\
&&\times\underline{B}^{*^i}({\bf
k},t)\underline{Y}^{*^j}_{\omega_1}({\bf
k}_1,t)\underline{Y}^{*^k}_{\omega_2}({\bf k}_2,t)+H.C]+...\nonumber\\
\end{eqnarray}
where $\int' d^3k$ denote the integration over the half space
$k_z\geq 0$ and $ \underline{{\bf E}}({\bf
k},t)=-\underline{\dot{{\bf A}}}({\bf k},t)-i{\bf
k}\underline{\varphi}({\bf k},t)$ and $\underline{{\bf B}}({\bf
k},t)=i{\bf k}\times\underline{{\bf A}}({\bf k},t)$.In the
Lagrangian (\ref{n9}) the three points ... indicate the other terms
which are dependent on the coupling tensors more than the third
rank. Hereafter we apply the symbols  $\int' d^3k$ and $\int d^3k$
for the integration over the half space$ k_z\geq 0$ and integration
over the total reciprocal space, respectively.\\
\section{Classical Euler-Lagrange equations}
The Lagrangian (\ref{n9}) does not involve the space derivatives of
the dynamical variables and one can easily obtain the classical
Euler-Lagrange equations for the Fourier transforms of the fields
appeared in the Lagrangian of the total system. The classical
Euler-Lagrange equation for the scalar potential
$\underline{\varphi}({\bf k},t)$  and vector potential $
\underline{{\bf A}}({\bf k},t)$ are respectively as follows
\begin{eqnarray}\label{n10}
&&\frac{d}{dt}(\frac{\delta L}{\delta
\underline{\dot{\varphi}}^*({\bf k},t)})-\frac{\delta L}{\delta
\underline{\varphi}^*({\bf k},t)}=0\nonumber\\
&&\Rightarrow -\varepsilon_0i {\bf k}\cdot\underline{\dot{{\bf
A}}}({\bf k},t)+\varepsilon_0 k^2\underline{\varphi}({\bf
k},t)=-i{\bf k}\cdot\underline{{\bf P}}({\bf k},t)
\end{eqnarray}
\begin{eqnarray}\label{n11}
&&\frac{d}{dt}(\frac{\delta L}{\delta \underline{\dot{A}}_i^*({\bf
k},t)})-\frac{\delta L}{\delta
\underline{A}_i^*({\bf k},t)}=0\hspace{2 cm}i=1,2,3\nonumber\\
&&\Rightarrow \mu_0 \varepsilon_0\underline{\ddot{{\bf A}}}({\bf
k},t)+\mu_0 \varepsilon_0 i{\bf k}\underline{\dot{\varphi}}({\bf
k},t)-{\bf k}\times({\bf k}\times\underline{{\bf A}}({\bf
k},t))\nonumber\\
&&=\mu_0\underline{\dot{{\bf P}}}({\bf k},t)+i\mu_0{\bf
k}\times\underline{{\bf M}}({\bf k},t)\nonumber\\
&&
\end{eqnarray}
where $ \underline{{\bf P}}({\bf k},t)$ and  $\underline{{\bf
M}}({\bf k},t)$ are the Fourier transforms of the electric and
magnetic polarization densities of the nonlinear magnetodielectric
medium, respectively. The components of the polarization densities
are defined in terms of the coupling tensors $\underline{{\bf
f}}^{(n)} , \underline{{\bf g}}^{(n)}$ and the the harmonic
oscillators modeling
 the medium as
\begin{eqnarray}\label{n12}
&&\underline{P}_i({\bf k},t)=\int_0^\infty d\omega_1 \int d^3k_1
\underline{f}^{(1)}_{ij}(\omega_1, {\bf k}, {\bf
k}_1)\underline{X}^j_{\omega_1}({\bf k}_1,t)\nonumber\\
&&+\int_0^\infty d\omega_1 \int_0^\infty d\omega_2\int d^3k_1 \int
d^3 k_2 \underline{f}^{(2)}_{ijk}(\omega_1,\omega_2,{\bf k}, {\bf
k}_1, {\bf k}_2)\underline{X}^j_{\omega_1}({\bf k}_1,t)
\underline{X}^k_{\omega_2}({\bf k}_2,t)\nonumber\\
&&+\int_0^\infty d\omega_1 \int_0^\infty d\omega_2 \int_0^\infty
d\omega_3\int d^3k_1 \int d^3 k_2 \int d^3k_3
\underline{f}^{(3)}_{ijkl}(\omega_1,\omega_2,\omega_3,{\bf k}, {\bf
k}_1, {\bf k}_2,{\bf k}_3)\nonumber\\
&&\times\underline{X}^j_{\omega_1}({\bf k}_1,t)
\underline{X}^k_{\omega_2}({\bf k}_2,t)\underline{X}^l_{\omega_3}({\bf k}_3,t)+....\nonumber\\
&&
\end{eqnarray}
\begin{eqnarray}\label{n13}
&&\underline{M}_i({\bf k},t)=\int_0^\infty d\omega_1 \int d^3k_1
\underline{g}^{(1)}_{ij}(\omega_1, {\bf k}, {\bf
k}_1)\underline{Y}^j_{\omega_1}({\bf k}_1,t)\nonumber\\
&&+\int_0^\infty d\omega_1 \int_0^\infty d\omega_2\int d^3k_1 \int
d^3 k_2 \underline{g}^{(2)}_{ijk}(\omega_1,\omega_2,{\bf k}, {\bf
k}_1, {\bf k}_2)\underline{Y}^j_{\omega_1}({\bf k}_1,t)
\underline{Y}^k_{\omega_2}({\bf k}_2,t)\nonumber\\
&&+\int_0^\infty d\omega_1 \int_0^\infty d\omega_2 \int_0^\infty
d\omega_3\int d^3k_1 \int d^3 k_2 \int d^3k_3
\underline{g}^{(3)}_{ijkl}(\omega_1,\omega_2,\omega_3,{\bf k}, {\bf
k}_1, {\bf k}_2,{\bf k}_3)\nonumber\\
&&\times\underline{Y}^j_{\omega_1}({\bf k}_1,t)
\underline{Y}^k_{\omega_2}({\bf k}_2,t)\underline{Y}^l_{\omega_3}({\bf k}_3,t)+....\nonumber\\
&&
\end{eqnarray}
\section{Canonical quantization}
In order to quantize the electromagnetic field canonically, one
should be eliminated the extra degrees of freedom from the total
Lagrangian (\ref{n9}). For this purpose we apply  the coulomb gauge
and eliminate the scalar potential $\underline{\varphi}({\bf k},t)$
from the total Lagrangian using the equation (\ref{n10}). In
reciprocal space the coulomb gauge is in the form ${\bf k
}\cdot\underline{{\bf A}}=0$. Applying the coulomb gauge and using
the equation (\ref{n10}) one can obtain the spatial Fourier
transform of the scalar potential as
\begin{equation}\label{n14}
\underline{\varphi}(\vec{k},t)=-\frac{\imath\vec{k}\cdot\underline{\vec{P}}(\vec{k},t)}{\varepsilon_0
k^2 }
\end{equation}
Let ${\bf e}_{{\bf k}\lambda}, \lambda=1,2$ and ${\bf e}_{{\bf
k}3}=\hat{k}=\frac{{\bf k}}{k}$ be three mutual orthogonal unit
vectors for each vector ${\bf k}$. Then from the coulomb gauge ${\bf
k }\cdot\underline{{\bf A}}=0 $, we are admissible expand the vector
potential along the orthogonal unit vector ${\bf e}_{{\bf
k}\lambda}, \lambda=1,2$ as
\begin{equation}\label{n15}
\underline{\vec{A}}(\vec{k},t)=\sum_{ \lambda=1}^2\
\underline{A}_\lambda( \vec{k},t)\vec{e}_{\lambda\vec{k}}
\end{equation}
where $\underline{A}_\lambda( \vec{k},t), \lambda=1,2$ are the new
dynamical variables of electromagnetic field. Also we introduce the
new dynamical variables $ \underline{X}_{\omega\lambda}({\bf k},t)$
and $ \underline{Y}_{\omega\lambda}({\bf k},t)$ for the nonlinear
magnetodielectric medium as the coefficients of the expansion of the
fields $ \underline{{\bf X}}_\omega({\bf k},t)$ and $
\underline{{\bf Y}}_\omega({\bf k},t)$ in terms of the orthogonal
unit vectors ${\bf e}_{{\bf k}\lambda}, \lambda=1,2,3$, that is
\begin{eqnarray}\label{n16}
\underline{\vec{X}}_\omega(\vec{k},t)&=&\sum_{ \lambda=1}^3
\underline{X}_{\omega\lambda}( \vec{k},t)\vec{e}_{\lambda\vec{k}}\nonumber\\
\underline{\vec{Y}}_\omega(\vec{k},t)&=&\sum_{ \lambda=1}^3
\underline{Y}_{\omega\lambda}( \vec{k},t)\vec{e}_{\lambda\vec{k}}
\end{eqnarray}
Substituting the scalar potential $ \underline{\varphi}({\bf k},t)$
from (\ref{n14}) in the Lagrangian (\ref{n9})and using the
expansions (\ref{n15}) and (\ref{n16}), one can rewrite the
Lagrangian of the total system in terms of the new coordinates of
the system as
\begin{eqnarray}\label{n17}
 L(t)&=&\int_0^\infty d\omega\ \int'd^3k
\sum_{\lambda=1}^3\left(|\dot{\underline{X}}_{\omega\lambda}|^2-\omega^2|\underline{X}_{\omega\lambda}|^2
 +|\dot{\underline{Y}}_{\omega\lambda}|^2-\omega^2|\underline{Y}_{\omega\lambda}|^2\right)\nonumber\\
& +& \int'd^3k
\sum_{\lambda=1}^2\left(\varepsilon_0|\dot{\underline{{A}}_\lambda}|^2
-\frac{ k^2|\underline{A}_{\lambda}|^2}{\mu_0} \right)
-\frac{1}{\varepsilon_0} \int' d^3k\
\frac{|{\bf k}\cdot\underline{{\bf P}}|^2}{k^2}\nonumber\\
& +&\int' d^3k \left[\left(-
\sum_{\lambda=1}^2\underline{\dot{A}}_\lambda {\bf e}_{\lambda {\bf
k}}\right)\cdot \underline{{\bf P}}^*({\bf k},t)+ \left( i{\bf
k}\times\sum_{\lambda=1}^2\underline{A}_\lambda {\bf e}_{\lambda
{\bf k}}\right)\cdot
\underline{{\bf M}}^*({\bf k},t)+H.c\right]\nonumber\\
&&
\end{eqnarray}
where the polarization densities $\underline{{\bf P}}({\bf k},t) ,
\underline{{\bf M}}({\bf k},t)$  have previously been defined in the
relations (\ref{n12}) and (\ref{n13}). The relation (\ref{n17}) give
us the Lagrangian of the total system in which the extra degrees of
freedom  have been eliminated and can be used for a canonical
quantization of the electromagnetic field together with the
nonlinear anisotropic magnetodielectric medium. The canonical
conjugate momenta of the total system for any vector ${\bf k}$ in
the half space $k_z\geq 0$ can be defined as
\begin{eqnarray}\label{n18}
-\underline{D}_\lambda (\vec{k},t)&=&\frac{\delta
\underline{L}}{\delta \left(\underline{\dot{A}}^*_\lambda
(\vec{k},t)\right)}=
\varepsilon_0\dot{\underline{A}}_\lambda ({\bf k},t)-{\bf e}_{\lambda {\bf k}}\cdot\underline{{\bf P}}(\vec{k},t)\nonumber\\
\underline{Q}_{\omega\lambda}(\vec{k},t)&=&\frac{\delta
\underline{L}}{\delta\left(
\underline{\dot{X}}^*_{\omega\lambda}(\vec{k},t)\right)}=
\underline{\dot{X}}_{\omega\lambda}(\vec{k},t)\nonumber\\
\underline{\Pi}_{\omega\lambda}(\vec{k},t)&=&\frac{\delta
\underline{L}}{\delta
\left(\underline{\dot{Y}}^*_{\omega\lambda}(\vec{k},t)\right)}=
\underline{\dot{Y}}_{\omega\lambda}(\vec{k},t)
\end{eqnarray}
Now, following the standard way for a canonical quantization,  we
impose the following commutation relations between the conjugate
dynamical variables of the total system
\begin{eqnarray}\label{n19}
\left[ \underline{A}^*_\lambda ( \vec{k},t) \ ,\
-\underline{D}_{\lambda'}( \vec{k'} ,
t)\right]&=&\imath\hbar\delta_{\lambda\lambda'}\delta(\vec{k}-\vec{k'})\nonumber\\
\left[ \underline{X}^*_{\omega\lambda }( \vec{k},t) \ ,\
\underline{Q}_{\omega'\lambda'}( \vec{k'} ,
t)\right]&=&\imath\hbar\delta_{\lambda\lambda'}\delta(\omega-\omega')\delta(\vec{k}-\vec{k'})\nonumber\\
\left[ \underline{Y}^*_{\omega\lambda }( \vec{k},t) \ ,\
\underline{\Pi}_{\omega'\lambda'}( \vec{k'} ,
t)\right]&=&\imath\hbar\delta_{\lambda\lambda'}\delta(\omega-\omega')\delta(\vec{k}-\vec{k'})\nonumber\\
&&
\end{eqnarray}
The Hamiltonian of the total system can be written as usual way as
\begin{eqnarray}\label{n20}
&&H(t)=\int'd^3 k\sum_{\lambda=1}^2\left[ -\underline{D}_\lambda
\underline{\dot{A}}^*_\lambda-\underline{D}^*_\lambda\underline{\dot{A}}_\lambda\right]\nonumber\\
&&+\int_0^\infty d\omega \int' d^3k\sum_{\lambda=1}^2\left[
\underline{Q}_{\omega\lambda}\underline{\dot{X}}^*_{\omega\lambda}+
\underline{Q}^*_{\omega\lambda}\underline{\dot{X}}_{\omega\lambda}\right]\nonumber\\
&&+\int_0^\infty d\omega \int' d^3k\sum_{\lambda=1}^2\left[
\underline{\Pi}_{\omega\lambda}\underline{\dot{Y}}^*_{\omega\lambda}+
\underline{\Pi}^*_{\omega\lambda}\underline{\dot{Y}}_{\omega\lambda}\right]-L(t)
\end{eqnarray}
Using the definitions of conjugate momenta given in (\ref{n18}), the
Hamiltonian of the system in terms of the coordinates of the system
and their conjugate momenta becomes
\begin{eqnarray}\label{n21}
H(t)&=&\int' d^3k \sum_{\lambda=1}^2\left(
\frac{|\underline{D}_{\lambda}-{\bf e}_{\lambda {\bf k}}\cdot
\underline{{\bf
P}}|^2}{\varepsilon_0}+\frac{k^2|\underline{A}_\lambda
|^2}{\mu_0}\right)+\frac{1}{\varepsilon_0}\int' d^3k\frac{|{\bf k}\cdot\underline{{\bf P}}|^2}{k^2}\nonumber\\
&-&\int' d^3k \left[\left( i{\bf
k}\times\sum_{\lambda=1}^2\underline{A}_{\lambda}{\bf
e}_{\lambda{\bf k}}\right)\cdot \underline{{\bf M}}^*+H.c\right]
\nonumber\\
&+&\int_0^\infty d\omega\int' d^3k\sum_{\lambda=1}^3\left( |\
\underline{Q}_{\omega \lambda}|^2+\omega^2|\ \underline{X}_{\omega
\lambda}|^2 + |\ \underline{\Pi}_{\omega \lambda}|^2+\omega^2|\
\underline{Y}_{\omega\lambda}|^2\right)\nonumber\\
&&
\end{eqnarray}
It can be easily shown that the Heisenberg equation for
$\underline{A}_\lambda({\bf k},t), \lambda=1,2$ leads to the
relation $\underline{{\bf D}}({\bf
k},t)=\varepsilon_0\underline{{\bf E}}({\bf k},t)+\underline{{\bf
P}}({\bf k},t)$ \cite{18}, where $\underline{{\bf
D}}=\sum_{\lambda=1}^2\underline{D}_\lambda {\bf e}_{{\bf
k}\lambda}$ is the displacement field and $\underline{{\bf
E}}=-\underline{\dot{{\bf A}}}-\frac{\hat{{\bf k}}(\hat{{\bf
k}}\cdot\underline{{\bf P}})}{\varepsilon_0}$ is the Fourier
transform of the electric field. Also the Heisenberg equation for
$\underline{D}_\lambda, \lambda=1,2$ give us the Maxwell equation
$\underline{\dot{{\bf D}}}({\bf k},t)=i{\bf k}\times\underline{{\bf
H}}({\bf k},t)$ in the reciprocal space , where $ \underline{{\bf
H}}=\frac{i{\bf k}\times\underline{{\bf A}}}{\mu_0}-\underline{{\bf
M}}$
is the magnetic induction field \cite{18}.\\
\section{The constitutive relations of the medium}
One can obtain the constitutive relations of the nonlinear
magnetodielectric medium using the definitions of polarization and
magnetization densities of the medium given by
(\ref{n12}),(\ref{n13}) and the Heisenberg equations for the
dynamical variables  $ \underline{X}_{\omega\lambda}({\bf k},t),
\underline{Y}_{\omega\lambda}({\bf k},t), \lambda=1,2,3$.
Straightforwardly, one can show that the combination of the
Heisenberg  equations of the conjugate variables $
\underline{X}_{\omega\lambda}({\bf k},t) ,
\underline{Q}_{\omega\lambda}({\bf k},t), \lambda=1,2,3$ leads to
the following equation of motion for the components of the field $
\underline{{\bf X}}({\bf k},t)$
\begin{eqnarray}\label{n22}
&& \underline{\ddot{X}}_{\omega i}({\bf
k},t)+\omega^2\underline{X}_{\omega i}({\bf k},t)=\int d^3k_1
\underline{f}^{\dag^{(1)}}_{ij}(\omega, {\bf k}_1, {\bf
k})\underline{E}^j({\bf k}_1,t)\nonumber\\
&&+\int_0^\infty d\omega_2\int d^3k_1 \int d^3k_2
\underline{f}^{*^(2)}_{jik}(\omega,\omega_2, {\bf k}_1, {\bf k},
{\bf k}_2)\underline{E}^j({\bf k}_1,t)\underline{X}^{*^k}({\bf
k}_2,t)\nonumber\\
&&+\int_0^\infty d\omega_1\int d^3k_1 \int d^3k_2
\underline{f}^{*^(2)}_{kji}(\omega_1,\omega, {\bf k}_2, {\bf k}_1,
{\bf k})\underline{E}^k({\bf k}_2,t)\underline{X}^{*^j}({\bf
k}_1,t)+...\nonumber\\
&&
\end{eqnarray}
where the completeness relation $ \sum_{\lambda=1}^3{\bf e}_{{\bf
k}\lambda i}{\bf e}_{{\bf k}\lambda j}=\delta_{ij}$ has been used ,
three points ... indicate the terms containing the coupling tensors
more than the third rank and $\underline{{\bf
E}}=-\underline{\dot{{\bf A}}}-\frac{\hat{{\bf k}}(\hat{{\bf
k}}\cdot\underline{{\bf P}})}{\varepsilon_0}$ is the Fourier
transform of the electric field. Similarly the combination of the
Heisenberg equations of the conjugate variables $
\underline{Y}_{\omega\lambda} , \underline{\Pi}_{\omega\lambda},
\lambda=1,2,3$ give us
\begin{eqnarray}\label{n23}
&& \underline{\ddot{Y}}_{\omega i}({\bf
k},t)+\omega^2\underline{Y}_{\omega i}({\bf k},t)=\int d^3k_1
\underline{g}^{\dag^{(1)}}_{ij}(\omega, {\bf k}_1, {\bf
k})\underline{B}^j({\bf k}_1,t)\nonumber\\
&&+\int_0^\infty d\omega_2\int d^3k_1 \int d^3k_2
\underline{g}^{*^(2)}_{jik}(\omega,\omega_2, {\bf k}_1, {\bf k},
{\bf k}_2)\underline{B}^j({\bf k}_1,t)\underline{Y}^{*^k}({\bf
k}_2,t)\nonumber\\
&&+\int_0^\infty d\omega_1\int d^3k_1 \int d^3k_2
\underline{g}^{*^(2)}_{kji}(\omega_1,\omega, {\bf k}_2, {\bf k}_1,
{\bf k})\underline{B}^k({\bf k}_2,t)\underline{Y}^{*^j}({\bf
k}_1,t)+...\nonumber\\
&&
\end{eqnarray}
where $ \underline{{\bf B}}({\bf k},t)=i{\bf k}\times\underline{{\bf
A}}({\bf k},t)$ is the Fourier transform of the magnetic field. In
(\ref{n22}) and (\ref{n23})the components of  the tensors $
\underline{{\bf f}}^{\dag^{(1)}} $ and $ \underline{{\bf
g}}^{\dag^{(1)}}$ are defined as $
\underline{f}^{\dag^{(1)}}_{ij}=\underline{f}^{*^{(1)}}_{ji} ,
\underline{g}^{\dag^{(1)}}_{ij}=\underline{g}^{*^{(1)}}_{ji}$The
differential equations (\ref{n22}) and (\ref{n23}) are two frames of
the complicated nonlinear coupled differential equations. The
solution of these equations exactly is impossible unless an
iteration method is used. One may apply the first order
approximation and keep only the first term in the right hand of the
equations (\ref{n22}) and (\ref{n23}) and write the solutions of the
differential equations (\ref{n22})and (\ref{n23}) as
\begin{eqnarray}\label{n24}
\underline{X}_{\omega i}({\bf k},t)=\underline{X}_{N\omega i}({\bf
k},t)+\int_{-\infty}^t dt' \frac{\sin \omega(t-t')}{\omega}\int
d^3k_1 \underline{f}^{\dag^{(1)}}_{ij}(\omega, {\bf k}_1, {\bf
k})\underline{E}^j({\bf k}_1,t')\nonumber\\
&&
\end{eqnarray}
\begin{eqnarray}\label{n25}
\underline{Y}_{\omega i}({\bf k},t)=\underline{Y}_{N\omega i}({\bf
k},t)+\int_{-\infty}^t dt' \frac{\sin \omega(t-t')}{\omega}\int
d^3k_1 \underline{g}^{\dag^{(1)}}_{ij}(\omega, {\bf k}_1, {\bf
k})\underline{B}^j({\bf k}_1,t')\nonumber\\
&&
\end{eqnarray}
where $\underline{{\bf X}}_{N\omega}({\bf k},t)$ and
$\underline{{\bf Y}}_{N\omega}({\bf k},t)$ are the solutions of the
homogenous equations
\begin{eqnarray}\label{n26}
&&\underline{\ddot{{\bf X}}}_{N\omega }({\bf
k},t)+\omega^2\underline{{\bf X}}_{N\omega }({\bf k},t)=0\nonumber\\
&&\underline{\ddot{{\bf Y}}}_{N\omega }({\bf
k},t)+\omega^2\underline{{\bf Y}}_{N\omega }({\bf k},t)=0
\end{eqnarray}
respectively which can be written as
\begin{eqnarray}\label{n27}
&&\underline{{\bf X}}_{N\omega}({\bf
k},t)=\sqrt{\frac{\hbar}{2\omega}}\sum_{\lambda=1}^3[b_\lambda(\omega,{\bf
k})e^{-i \omega t}+b^\dag_\lambda(\omega,{\bf k})e^{i \omega t}]{\bf
e}_{{\bf k}\lambda}\nonumber\\
&&\underline{{\bf Y}}_{N\omega}({\bf
k},t)=\sqrt{\frac{\hbar}{2\omega}}\sum_{\lambda=1}^3[d_\lambda(\omega,{\bf
k})e^{-i \omega t}+d^\dag_\lambda(\omega,{\bf k})e^{i \omega t}]{\bf
e}_{{\bf k}\lambda}
\end{eqnarray}
where the operators $b_\lambda(\omega,{\bf k}),
b^\dag_\lambda(\omega,{\bf k})$ and $d_\lambda(\omega,{\bf k}),
d^\dag_\lambda(\omega,{\bf k})$ satisfy the commutation relations
\begin{eqnarray}\label{n28}
&&[b_\lambda(\omega,{\bf k}), b^\dag_{\lambda'}(\omega',{\bf
k'})]=\delta_{\lambda\lambda'}\delta(\omega-\omega')\delta({\bf
k}-{\bf k'})\nonumber\\
&&[d_\lambda(\omega,{\bf k}), d^\dag_{\lambda'}(\omega',{\bf
k'})]=\delta_{\lambda\lambda'}\delta(\omega-\omega')\delta({\bf
k}-{\bf k'})
\end{eqnarray}
in compatible to the commutation relations (\ref{n19}).\\
Now substituting $\underline{{\bf X}}_\omega({\bf k},t)$ from
(\ref{n24}) in (\ref{n12}) and $\underline{{\bf Y}}_\omega({\bf
k},t)$ from (\ref{n25}) in (\ref{n13}) give us the constitutive
relations of the nonlinear anisotropic magnetodielectric medium as
\begin{eqnarray}\label{n29}
&&\underline{P}_i({\bf k},t)=\underline{P}_{Ni}({\bf
k},t)+\int_{-\infty}^{+\infty}dt_1\int d^3k_1
\chi^{{(1)}}_{ij}(t-t_1, {\bf k}, {\bf k}_1)\underline{E}^j({\bf
k}_1,t_1)\nonumber\\
&&+\int_{-\infty}^{+\infty}dt_1\int_{-\infty}^{+\infty}dt_2\int
d^3k_1\int d^3k_2 \chi^{{(2)}}_{ijk}(t-t_1, t-t_2, {\bf k}, {\bf
k}_1, {\bf k}_2)\underline{E}^j({\bf k}_1,t_1)\underline{E}^k({\bf
k_2},t_2)\nonumber\\
&&+......
\end{eqnarray}
\begin{eqnarray}\label{n30}
&&\underline{M}_i({\bf k},t)=\underline{M}_{Ni}({\bf
k},t)+\int_{-\infty}^{+\infty}dt_1\int d^3k_1
\zeta^{{(1)}}_{ij}(t-t_1, {\bf k}, {\bf k}_1)\underline{B}^j({\bf
k}_1,t_1)\nonumber\\
&&+\int_{-\infty}^{+\infty}dt_1\int_{-\infty}^{+\infty}dt_2\int
d^3k_1\int d^3k_2 \zeta^{{(2)}}_{ijk}(t-t_1, t-t_2, {\bf k}, {\bf
k}_1, {\bf k}_2)\underline{B}^j({\bf k}_1,t_1)\underline{B}^k({\bf
k_2},t_2)\nonumber\\
&&+......
\end{eqnarray}
where the tensors $\chi^n$ and $\zeta^n$ for $ n=1,2,... $ are
respectively the electric and magnetic susceptibility tensors and
define in terms of the coupling tensors $\underline{{\bf f}}^{(n)}$
and $\underline{{\bf g}}^{(n)}$ as
\begin{eqnarray}\label{n31}
&&\chi^{(n)}_{ii_1i_2...i_n}(t_1,t_2,...,t_n,{\bf k},{\bf k}_1,{\bf
k}_2,...,{\bf k}_n)=\nonumber\\
&&\Theta(t_1)\Theta(t_2)....\Theta(t_n)\int_0^\infty
d\omega_1\int_0^ d\omega_2....\int_0^\infty d\omega_n \frac{\sin
\omega_1 t_1}{\omega_1}\frac{\sin \omega_2
t_2}{\omega_2}....\frac{\sin \omega_n t_n}{\omega_n}\nonumber\\
&&\times\int d^3p_1\int d^3p_2....\int d^3p_n
 [\underline{f}^{(n)}_{ij_1j_2...j_n}(\omega_1,\omega_2,...,\omega_n,{\bf
 k},{\bf p}_1,{\bf p}_2,...{\bf p}_n)\nonumber\\
 &&\times \underline{f}^{\dag^{(1)}}_{j_1i_1}(\omega_1,{\bf k}_1,{\bf
 p}_1) \underline{f}^{\dag^{(1)}}_{j_2i_2}(\omega_2,{\bf k}_2,{\bf
 p}_2)....\underline{f}^{\dag^{(1)}}_{j_ni_n}(\omega_n,{\bf k}_n,{\bf
 p}_n)]
\end{eqnarray}
\begin{eqnarray}\label{n32}
&&\zeta^{(n)}_{ii_1i_2...i_n}(t_1,t_2,...,t_n,{\bf k},{\bf k}_1,{\bf
k}_2,...,{\bf k}_n)=\nonumber\\
&&\Theta(t_1)\Theta(t_2)....\Theta(t_n)\int_0^\infty
d\omega_1\int_0^ d\omega_2....\int_0^\infty d\omega_n \frac{\sin
\omega_1 t_1}{\omega_1}\frac{\sin \omega_2
t_2}{\omega_2}....\frac{\sin \omega_n t_n}{\omega_n}\nonumber\\
&&\times\int d^3p_1\int d^3p_2....\int d^3p_n
 [\underline{g}^{(n)}_{ij_1j_2...j_n}(\omega_1,\omega_2,...,\omega_n,{\bf
 k},{\bf p}_1,{\bf p}_2,...{\bf p}_n)\nonumber\\
 &&\times \underline{g}^{\dag^{(1)}}_{j_1i_1}(\omega_1,{\bf k}_1,{\bf
 p}_1) \underline{g}^{\dag^{(1)}}_{j_2i_2}(\omega_2,{\bf k}_2,{\bf
 p}_2)....\underline{g}^{\dag^{(1)}}_{j_ni_n}(\omega_n,{\bf k}_n,{\bf
 p}_n)]
\end{eqnarray}
where $\Theta(t)$ is the step function and over the repeated indices
$j_1,j_2,...j_n $ should be summed. There are some symmetry
conditions that the electric and magnetic susceptibility tensors of
various rank should satisfy.The tensors $ \chi^{(n)}$ and $
\zeta^{(n)}$ may satisfy the symmetry conditions \cite{20.1}
\begin{eqnarray}\label{n33}
&&\chi^{(n)}_{ii_1i_2...i_\alpha...i_\beta...i_n}(t_1,t_2,...,t_\alpha,...,t_\beta,...,t_n,{\bf
k},{\bf k}_1,{\bf k}_2,...,{\bf k}_\alpha,...,{\bf k}_\beta,...,{\bf
k}_n)\nonumber\\
&&=\chi^{(n)}_{ii_1i_2...i_\beta...i_\alpha...i_n}(t_1,t_2,...,t_\beta,...,t_\alpha,...,t_n,{\bf
k},{\bf k}_1,{\bf k}_2,...,{\bf k}_\beta,...,{\bf k}_\alpha,...,{\bf
k}_n)\nonumber\\
\end{eqnarray}
\begin{eqnarray}\label{n34}
&&\zeta^{(n)}_{ii_1i_2...i_\alpha...i_\beta...i_n}(t_1,t_2,...,t_\alpha,...,t_\beta,...,t_n,{\bf
k},{\bf k}_1,{\bf k}_2,...,{\bf k}_\alpha,...,{\bf k}_\beta,...,{\bf
k}_n)\nonumber\\
&&=\zeta^{(n)}_{ii_1i_2...i_\beta...i_\alpha...i_n}(t_1,t_2,...,t_\beta,...,t_\alpha,...,t_n,{\bf
k},{\bf k}_1,{\bf k}_2,...,{\bf k}_\beta,...,{\bf k}_\alpha,...,{\bf
k}_n)\nonumber\\
\end{eqnarray}
From the definitions (\ref{n31}) and (\ref{n32}) it is clear that
the symmetry relations (\ref{n33}) and (\ref{n34}) are satisfied
provided that we impose the conditions
\begin{eqnarray}\label{n35}
&&\underline{f}^{(n)}_{ij_1j_2...j_\alpha...j_\beta...j_n}(\omega_1,\omega_2,...,\omega_\alpha,...,\omega_\beta,...,\omega_n
,{\bf k},{\bf k}_1,{\bf k}_2,...,{\bf k}_\alpha,...,{\bf
k}_\beta,...,{\bf k}_n)\nonumber\\
&&=\underline{f}^{(n)}_{ij_1j_2...j_\beta...j_\alpha...j_n}(\omega_1,\omega_2,...,\omega_\beta,...,\omega_\alpha,...,\omega_n
,{\bf k},{\bf k}_1,{\bf k}_2,...,{\bf k}_\beta,...,{\bf
k}_\alpha,...,{\bf k}_n)\nonumber\\
\end{eqnarray}
\begin{eqnarray}\label{n35.1}
&&\underline{g}^{(n)}_{ij_1j_2...j_\alpha...j_\beta...j_n}(\omega_1,\omega_2,...,\omega_\alpha,...,\omega_\beta,...,\omega_n
,{\bf k},{\bf k}_1,{\bf k}_2,...,{\bf k}_\alpha,...,{\bf
k}_\beta,...,{\bf k}_n)\nonumber\\
&&=\underline{g}^{(n)}_{ij_1j_2...j_\beta...j_\alpha...j_n}(\omega_1,\omega_2,...,\omega_\beta,...,\omega_\alpha,...,\omega_n
,{\bf k},{\bf k}_1,{\bf k}_2,...,{\bf k}_\beta,...,{\bf
k}_\alpha,...,{\bf k}_n)\nonumber\\
\end{eqnarray}
on the coupling tensors $ \underline{{\bf f}}^{(n)}$ and $
\underline{{\bf g}}^{(n)}$ for $ n=2,3,...$.\\
In constitutive relations (\ref{n29}) and (\ref{n30}) $
\underline{{\bf P}}_N({\bf k},t)$ and $ \underline{{\bf M}}_N({\bf
k},t)$ are the noise polarization and magnetization fields,
respectively, where their components are written in terms of the
$\underline{{\bf X}}_{N\omega}({\bf k},t)$ and $\underline{{\bf
Y}}_{N\omega}({\bf k},t)$  as
\begin{eqnarray}\label{n36}
&&\underline{P}_{Ni}({\bf k},t)=\int_0^\infty d\omega_1\int d^3k_1
\underline{f}^{(1)}_{ij}(\omega_1,{\bf k},{\bf
k}_1)\underline{X}_{N\omega_1}^j({\bf k}_1,t)\nonumber\\
&&+\int_0^\infty d\omega_1\int_0^\infty d\omega_2\int d^3k_1\int d^3
k_2 [\underline{f}^{(2)}_{ijk}(\omega_1,\omega_2,{\bf k},{\bf
k}_1,{\bf k}_2)\underline{X}_{N\omega_1}^j({\bf
k}_1,t)\underline{X}_{N\omega_2}^k({\bf k}_2,t)]\nonumber\\
&& \int_0^\infty d\omega_1\int_0^\infty d\omega_2\int d^3k_1\int d^3
k_2 \left[\underline{f}^{(2)}_{ijk}(\omega_1,\omega_2,{\bf k},{\bf
k}_1,{\bf k}_2) \int_0^t
dt'\frac{\sin\omega_2(t-t')}{\omega_2}\right.\nonumber\\
&&\times\left. \int
d^3p_2\underline{f}^{\dag^{(1)}}_{kl}(\omega_2,{\bf p}_2,{\bf k}_2)
\left(\underline{X}_{N\omega_1}^j({\bf k}_1,t)\underline{E}^l({\bf
p}_2,t')+\underline{E}^l({\bf p}_2,t')\underline{X}^j_{N\omega_1}({\bf k}_1,t)\right)\right]+...\nonumber\\
&&
\end{eqnarray}
\begin{eqnarray}\label{n37}
&&\underline{M}_{Ni}({\bf k},t)=\int_0^\infty d\omega_1\int d^3k_1
\underline{g}^{(1)}_{ij}(\omega_1,{\bf k},{\bf
k}_1)\underline{Y}_{N\omega_1}^j({\bf k}_1,t)\nonumber\\
&&+\int_0^\infty d\omega_1\int_0^\infty d\omega_2\int d^3k_1\int d^3
k_2 [\underline{g}^{(2)}_{ijk}(\omega_1,\omega_2,{\bf k},{\bf
k}_1,{\bf k}_2)\underline{Y}_{N\omega_1}^j({\bf
k}_1,t)\underline{Y}_{N\omega_2}^k({\bf k}_2,t)]\nonumber\\
&& \int_0^\infty d\omega_1\int_0^\infty d\omega_2\int d^3k_1\int d^3
k_2 \left[\underline{g}^{(2)}_{ijk}(\omega_1,\omega_2,{\bf k},{\bf
k}_1,{\bf k}_2) \int_0^t
dt'\frac{\sin\omega_2(t-t')}{\omega_2}\right.\nonumber\\
&&\times\left. \int
d^3p_2\underline{g}^{\dag^{(1)}}_{kl}(\omega_2,{\bf p}_2,{\bf k}_2)
\left(\underline{Y}_{N\omega_1}^j({\bf k}_1,t)\underline{B}^l({\bf
p}_2,t')+\underline{B}^l({\bf p}_2,t')\underline{Y}^j_{N\omega_1}({\bf k}_1,t)\right)\right]+...\nonumber\\
&&
\end{eqnarray}
where the symmetry conditions (\ref{n35}) and (\ref{n35.1}) have been used ,
 the summation should be done on the repeated indices and and the three
points ... indicate the terms containing the coupling tensors more
than the third rank.\\
\section{The time dependence of the electric and magnetic field}
 Let we
write the the electric field  $ \underline{{\bf E}}({\bf k},t)$ in
terms  of its temporal Fourier transform  as
\begin{equation}\label{n38}
\underline{{\bf E}}({\bf k},t)=\frac{1}{(2\pi)^{\frac{3}{2}}}\int
_{-\infty}^{+\infty} d\omega \underline{\tilde{{\bf E}}}({\bf
k},\omega)e^{i\omega t}
\end{equation}
then, using this relation and similar transformations for the other
fields and combination of the Maxwell equations $i{\bf k}\times {\bf
E}({\bf k},t)=\dot{\underline{{\bf B}}}({\bf k},t) ,
\dot{\underline{{\bf D}}}({\bf k},t)=i{\bf k}\times\underline{{\bf
H}}({\bf k},t)$ give us
\begin{equation}\label{n39}
{\bf k}\times{\bf k}\times \underline{\tilde{{E}}}({\bf
k},\omega)+\frac{\omega^2}{c^2}\underline{\tilde{{\bf E}}}({\bf
k},\omega)= -\mu_0\omega^2 \underline{\tilde{{\bf P}}}({\bf
k},\omega)-\mu_0\omega{\bf k}\times\underline{\tilde{{\bf M}}}({\bf
k},\omega)
\end{equation}
for the electric field in frequency domain where
$\underline{\tilde{{\bf P}}}({\bf k},\omega)$ and
$\underline{\tilde{{\bf M}}}({\bf k},\omega)$ are the temporal
Fourier transforms of $\underline{{\bf P}}({\bf k},t)$ and
$\underline{{\bf M}}({\bf k},t)$, respectively. Now substituting
$\underline{\tilde{{\bf P}}}({\bf k},\omega)$ and $
\underline{\tilde{{\bf M}}}({\bf k},\omega)$, using the constitutive
relations (\ref{n29}) and (\ref{n30}), in (\ref{n39}) leads to the
integral equation
\begin{eqnarray}\label{n40}
&&\Lambda_{ii_1}({\bf k},\omega)\underline{\tilde{E}}_{i_1}({\bf
k},\omega) +(2\pi)^{\frac{3}{2}}\mu_0\omega^2\int
d^3k_1\tilde{\chi}^{(1)}_{ii_1}(\omega,{\bf k},{\bf
k}_1)\tilde{E}_{i_1}({\bf k}_1,\omega)\nonumber\\
&&+(2\pi)^{\frac{3}{2}}\mu_0\omega^2\int d^3k_1\int
d^3k_2\int_0^\infty
d\omega_1\nonumber\\
&&\times[\tilde{\chi}^{(2)}_{ii_1i_2}(\omega_1,\omega-\omega_1,{\bf
k},{\bf k}_1,{\bf k}_2)\tilde{E}_{i_1}({\bf
k}_1,\omega_1)\tilde{E}_{i_2}({\bf
k}_2,\omega-\omega_1)]\nonumber\\
&&+......\nonumber\\
&&+(2\pi)^{\frac{3}{2}}\mu_0\omega\int
d^3k_1\epsilon_{ij_1j_2}k_{j_1}\tilde{\zeta}^{(1)}_{j_2i_1}(\omega,{\bf
k},{\bf
k}_1)\frac{({\bf k}_1\times\tilde{E}({\bf k}_1,\omega))_{i_1}}{-\omega}\nonumber\\
&&+(2\pi)^{\frac{3}{2}}\mu_0\omega\int d^3k_1\int
d^3k_2\int_0^\infty d\omega_1\nonumber\\
&&\times\left[\epsilon_{ij_1j_2}k_{j_1}\tilde{\zeta}^{(2)}_{j_2i_1i_2}(\omega_1,\omega-\omega_1,{\bf
k},{\bf k}_1,{\bf k}_2)\right.\nonumber\\
&&\times\left.\frac{({\bf k}_1\times\tilde{E}({\bf
k}_1,\omega_1))_{i_1}}{-\omega_1}\frac{({\bf
k}_2\times\tilde{E}({\bf
k}_2,\omega-\omega_1))_{i_2}}{-(\omega-\omega_1)}\right]+...
=\underline{\tilde{J}}_i({\bf k},\omega)
\end{eqnarray}
for the electric field where the Maxwell equation $ i{\bf
k}\times\underline{{\bf E}}({\bf k},t)=\underline{\dot{{\bf
B}}}({\bf k},t)$  has been used, $\epsilon_{ijk}$ is three
dimensional levi-Civita symbol and
\begin{eqnarray}\label{n41}
&&\Lambda_{ii_1}({\bf
k},\omega)=\epsilon_{ijm}\epsilon_{mni_1}k_jk_n+\frac{\omega^2}{c^2}\nonumber\\
&&\underline{\tilde{J}}_i({\bf
k},\omega)=-\mu_0\omega^2\underline{\tilde{P}}_{Ni}({\bf k}
,\omega)-\mu_0\omega\epsilon_{ijl}k_j\underline{\tilde{M}}_{Nl}({\bf
k} ,\omega)\nonumber\\
&&
\end{eqnarray}
In integral equation (\ref{n40}) $ \tilde{\chi}^{(n)},
\tilde{\zeta}^{(n)}, n=1,2,...$ are the temporal Fourier transforms
of the Susceptibility tensors $\chi^{(n)}, \zeta^{(n)}$ which is
defined as
\begin{eqnarray}\label{n42}
&&\chi^{(n)}_{ii_1i_2...i_n}(t_1,t_2,...,t_n,{\bf k},{\bf
k}_1,...,{\bf k}_n)
=\frac{1}{(2\pi)^{\frac{3(n+1)}{2}}}\int_{-\infty}^{+\infty}d\omega_1\int_{-\infty}^{+\infty}d\omega_2
...\int_{-\infty}^{+\infty}d\omega_n\nonumber\\
&&\times[
\chi^{(n)}_{ii_1...i_n}(\omega_1,\omega_2,...,\omega_n,{\bf
k}_1,{\bf k}_2,...,{\bf
k}_n)e^{i\omega_1t_1+i\omega_2t_2+...i\omega_nt_n}]\nonumber\\
&&
\end{eqnarray}
and $\tilde{\zeta}^{(n)}$ is defined in a similar way.\\
The equation (\ref{n40}) is an integral equation with kerenels $
\tilde{\chi}^{(n)}$ and $\tilde{\zeta}^{(n)}$ and the source term
$\underline{\tilde{{\bf J}}}({\bf k},\omega)$. As well, from the
definitions of the noise polarization and magnetization densities
given by (\ref{n36}) and(\ref{n37}), it is seen that the source term
$\underline{\tilde{{\bf J}}}({\bf k},\omega)$ is dependent on the
electric and magnetic fields. The integral equations such as
(\ref{n40}) can be solved by an iteration method\cite{21}. In the
zero order approximation we neglect the integrals in the left hand
of the equation(\ref{n40}) and write the components of the electric
field $\underline{\tilde{{\bf E}}}({\bf k},\omega) $ as follows
\begin{equation}\label{n43}
\underline{\tilde{E}}^{(0)}_i({\bf k},\omega)=\Lambda^{-1}_{ij}({\bf
k},\omega)\underline{\tilde{J}}^{(0)}_j({\bf k},\omega)
\end{equation}
where $\underline{\tilde{{\bf J}}}^{(0)}({\bf k},\omega)$ is the
source term $\underline{\tilde{{\bf J}}}({\bf k},\omega)$ given by
(\ref{n41}) from which the terms containing the electric and
magnetic fields have been eliminated. Therefore in the zero order
approximation the time dependence of the electric and magnetic
fields can be written as
\begin{eqnarray}\label{n44}
&&\underline{{\bf E}}^{(0)}({\bf
k},t)=\frac{1}{(2\pi)^{\frac{3}{2}}}\int_{-\infty}^{+\infty} d\omega
\underline{\tilde{{\bf E}}}^{(0)}({\bf k},\omega) e^{i\omega
t}\nonumber\\
&&\underline{{\bf B}}^{(0)}({\bf
k},t)=\frac{1}{(2\pi)^{\frac{3}{2}}}\int_{-\infty}^{+\infty} d\omega
\frac{{\bf k}\times\underline{\tilde{{\bf E}}}^{(0)}({\bf
k},\omega)}{-\omega} e^{i\omega t}
\end{eqnarray}
In $n^,$th  order approximation the components of the temporal
Fourier transform of the electric field for $n=1,2,3,...$ can be
computed using the following recurrence relation
\begin{eqnarray}\label{n45}
&&\underline{\tilde{E}}^{(n)}_i({\bf k},\omega)=
-(2\pi)^{\frac{3}{2}}\mu_0\omega^2\int d^3k_1
\left[\Lambda^{(-1)}_{ij}({\bf
k},\omega)\tilde{\chi}^{(1)}_{ji_1}(\omega,{\bf k},{\bf
k}_1)\tilde{E}^{(n-1)}_{i_1}({\bf k}_1,\omega)\right]\nonumber\\
&&-(2\pi)^{\frac{3}{2}}\mu_0\omega^2\int d^3k_1\int
d^3k_2\int_0^\infty d\omega_1\nonumber\\
&& \times\left[\Lambda^{(-1)}_{ij}({\bf k},\omega)
\tilde{\chi}^{(2)}_{ji_1i_2}(\omega_1,\omega-\omega_1,{\bf k},{\bf
k}_1,{\bf k}_2)\tilde{E}^{(n-1)}_{i_1}({\bf k}_1,\omega_1)
\tilde{E}^{(n-1)}_{i_2}({\bf k}_2,\omega-\omega_1)\right]+...\nonumber\\
&&-(2\pi)^{\frac{3}{2}}\mu_0\omega\int
d^3k_1\left[\Lambda^{(-1)}_{ij}({\bf
k},\omega)\epsilon_{jj_1j_2}k_{j_1}\tilde{\zeta}^{(1)}_{j_2i_1}(\omega,{\bf
k},{\bf k}_1)
\frac{[{\bf k}_1\times\tilde{E}^{(n-1)}({\bf k}_1,\omega)]_{i_1}}{-\omega}\right]\nonumber\\
&&-(2\pi)^{\frac{3}{2}}\mu_0\omega\int d^3k_1\int
d^3k_2\int_0^\infty d\omega_1\nonumber\\
&&\times \left[\Lambda^{(-1)}_{ij}({\bf k},\omega)
\epsilon_{jj_1j_2}k_{j_1}\tilde{\zeta}^{(2)}_{j_2i_1i_2}(\omega_1,\omega-\omega_1,{\bf
k},{\bf
k}_1,{\bf k}_2)\right.\nonumber\\
&&\times\left.\frac{[{\bf k}_1\times\tilde{E}^{(n-1)}({\bf
k}_1,\omega_1)]_{i_1}}{-\omega_1}\frac{[{\bf
k}_2\times\tilde{E}^{(n-1)}({\bf
k}_2,\omega-\omega_1)]_{i_2}}{-(\omega-\omega_1)}\right] +...
\nonumber\\
&&+\Lambda^{(-1)}_{ij}({\bf
k},\omega)\underline{\tilde{J}}^{(n-1)}_j({\bf k},\omega)
\end{eqnarray}
where the three points ...indicate the terms containing the electric
and magnetic susceptibility tensors more than the third rank and the
components of $\underline{\tilde{{\bf J}}}^{(n-1)}({\bf k},\omega)$
are defined as
\begin{eqnarray}\label{n46}
&&\underline{\tilde{J}}^{(n-1)}_i({\bf
k},\omega)=-\mu_0\omega^2\underline{\tilde{P}}^{(n-1)}_{Ni}({\bf k}
,\omega)-\mu_0\omega\epsilon_{ijl}k_j\underline{\tilde{M}}^{(n-1)}_{Nl}({\bf
k} ,\omega)
\end{eqnarray}
and $\underline{\tilde{{\bf P}}}_N^{(n-1)}({\bf k},\omega)$ and
$\underline{{\tilde{\bf M}}}_N^{(n-1)}({\bf k},\omega)$ are the
temporal Fourier transforms of the noise polarization and
magnetization fields given by (\ref{n36}) and (\ref{n37}) in which
the electric and magnetic fields $\underline{{\bf E}}^{(n-1)}({\bf
k},t) , \underline{{\bf B}}^{(n-1)}({\bf k},t)$ are inserted in
those terms of $ \underline{{\bf P}}_N({\bf k},t) , \underline{{\bf
M}}_N({\bf k},t)$ that are dependent on the electric and magnetic
fields.\\
\section{Conclusion}
By introducing a Lagrangian for the total system , that is the
nonlinear anisotropic magnetodielectric medium plus the
electromagnetic field, a canonical quantization was introduced for
both the medium and the electromagnetic field. The coupling tensors
appeared in the Lagrangian of the total system had a crucial role in
this quantization scheme. So that the electric and magnetic
polarization fields could be written in terms of the coupling
tensors and the space dependent three dimensional harmonic
oscillators modeling the medium. As well the electric and magnetic
susceptibility tensors of the medium were obtained in terms of the
coupling tensors. Also the constitutive relations of the
magnetodielectric medium successfully could be obtained using the
Heisenberg equations of the harmonic oscillators modeling the
medium.


\begin{thebibliography}{99}
\bibitem{1}  H. B. G. Casimir, Proc. K. Ned. Akad. Wet.51,793(1948).
\bibitem{2}  P. W. Milonni, \emph{The quantum vacuum: An introduction to quantum electrodynamics} (Academic Press, San Diego,1994).
\bibitem{3}  H. Casimir, D. Polder, Phys. Rev. A 73, 360(1948).
\bibitem{4}  E. Lifshitz, \emph{The theory of molecular attractive forces betweensolids}(1956).
\bibitem{5}  I. E. e. Dzyaloshinskii, E. Lifshitz, L. P. Pitaevskii,\emph{General theory of Van der waals forces}, Physics-Uspekhi 4,153-176(1961).
\bibitem{6}  C. Raabe, D. G. Welsch, Phys. Rev. A 71, 013814 (2005).
\bibitem{7}  R. Matloob, Phys. Rev. A 70, 062110(2004).
\bibitem{8}  R. Matloob, Phys. Rev. A 60, 3421(1999).
\bibitem{9}  R. Podgornik, P. L. Hansen, V. A. Parsegian, J. Chem. Phys. Vol. 119, No. 2(2003).
\bibitem{10} R. Golestanian, Phys. Rev. Lett. 95, 230601 (2005).
\bibitem{11}  E. M. purcell, Phys. Rev. 69, 681 (1946).
\bibitem{12}  E. Yablonovitch, Phys. Rev. Lett. 58,2059(1987).
\bibitem{13}  S. M. Barnett, B. Huttner, R. Loudon, Phys. Rev. Lett.68,3698(1992).
\bibitem{14}  S. M. Barnett, B. Huttner, R. Loudon, R. Matloob, J. Phys. B 29, 3763(1996).
\bibitem{15}  B. Huttner, S. M. Barnett, Phys. Rev. A 46, 4306(1992)
\bibitem{16}  H. T. Dung, L. Kn\"{o}ll, D. G. Welsch, Phys. Rev. A 57, 3931(1998).
\bibitem{17}  F. Kheirandish, M. Amooshahi, M. Soltani, J. Phys. B: At. Mol. Opt. Phys. 42, 075504(2009).
\bibitem{18}  M. Amooshahi, J. Math. Phys. 50, 062301(2009)
\bibitem{19} M. Amooshahi, Eur. Phys. J. D 69 (2015)
\bibitem{20}  M. Amooshahi, E. Amooghorban, Annals of physics, 325, 1976-1986 (2010)
\bibitem{20.1} Guang S. He, Song H. Liu, \emph{Physics of nonlinear
optics}, World Scientific (1999).
\bibitem{21}  Sadri Hassani, \emph{Foundations of Mathematical Physics}, McGraw-Hill )1991)
\end{thebibliography}
\end{document}